\documentstyle[referee]{laa}

\newcommand {\apj} {ApJ}
\newcommand {\apjs} {ApJS}
\newcommand {\aj} {AJ}

\newcommand {\aaa}  {\mbox{A\&A}}
\newcommand {\aaas} {\mbox{A\&AS}}
\newcommand {\pasp} {PASP}

\newcommand {\li} {Li\,{\sc I}\,$\lambda$670.8~nm}
\newcommand {\lii} {Li\,{\sc I}}

\def\et{ et al.~}
\def\ha{\relax \ifmmode {\rm H}\alpha\else H$\alpha$\fi}
\def\hb{\relax \ifmmode {\rm H}\beta\else H$\beta$\fi}
\def\deg{\hbox{$^\circ$}}
\def\min{\hbox{$^\prime$}}
\def\sec{\hbox{$^{\prime\prime}$}}

\def\x{$\pm$}
\def\lee{$\le$}
\def\lt{$<$}

\def\kms{km s$^{-1}$}

\begin{document}

\thesaurus{06(02.03.2; 08.01.1; 08.12.1; 08.16.5; 08.18.1)}

\title{Pre-main sequence lithium burning}

\subtitle {I. Weak  T~Tauri stars\thanks{Based on observations made with the
William Herschel, Isaac Newton and Jacobus Kapteyn telescopes, operated on the
island of La~Palma by the Royal Greenwich Observatory in the Spanish
Observatorio del Roque de los Muchachos of the Instituto de Astrof\'\i sica de
Canarias.}}

\author{E.L.~Mart\'\i n\inst{1} \and R.~Rebolo\inst{1} \and
A.~Magazz\`u\inst{2}
\and Ya.V.~Pavlenko\inst{3}}

\institute {Instituto de Astrof\'\i sica de Canarias, E-38200 La
Laguna, Tenerife, Spain
\and
Osservatorio Astrofisico di Catania, I-95125 Catania, Italy
\and
The Principal Observatory of Ukraine, Kiev, Ukraine}

\date{received June 29, 1993; accepted July 20, 1993}

\maketitle

\begin{abstract}
We report high-resolution spectroscopic observations  for
a sample of 38~T Tauri stars (TTS),
complemented by  UBVRI photometry of 13~TTS, and CCD VRI photometry for 2
visual binaries.
Based on these observations
and data taken from the literature, we derive lithium abundances
in 53 TTS, concentrating on weak-line TTS (WTTS). The sample spans
the range in spectral types from K0-M3,
aproximately corresponding to masses between 1.2 and 0.2~M$_\odot$.

Our study of the statistical distribution of lithium abundances in
WTTS gives the following results:
(1) At luminosities $\ge$~0.9~L$_\odot$ the Li abundances are remarkably
uniform.  The mean value, log~N(Li)=3.1, coincides with the ``cosmic" lithium
abundance.
(2) We find strong evidence for PMS lithium burning.
Significant Li depletion appears below 0.5~L$_\odot$
in the mass range 0.9-0.2~M$_\odot$ and increases towards lower luminosities.
Current theoretical evolutionary models do not seem to fit consistently
the observed pattern of Li abundances in the whole mass range. In particular,
at the lower mass end (0.4-0.2~M$_\odot$), the observed luminosity of the
Li burning turning point is about a factor 4 higher than
predicted by the models.
At masses 1.2-1.0~M$_\odot$ the observations imply less PMS Li burning than
theoretically expected.

Investigation of a possible relation between lithium and
rotation in TTS shows that: (1)
Low Li abundances appear only among stars with low v~sin{\it i}.
Fast rotators with masses around 0.8~M$_\odot$ do not show evidence for strong
Li depletion towards lower luminosities as slow rotators do.
(2) In a sample restricted to only K5-K7 stars with photometrically measured
rotational periods,  we find that the angular momentum spread before
Li burning begins is larger than a factor 10.

Lithium depletion associated to angular momentum loss during PMS evolution is
not required to explain the observed abundances. The observations suggest
that the efficiency of PMS Li burning in the mass range  0.9-0.7~M$_\odot$
is reduced in the presence of rapid rotation.
\end{abstract}

\section{Introduction}
\label{sec1}

The presence of the \lii\ resonance line has usually been a useful criterion
for identifying low-mass pre-main sequence (PMS) stars (Walter \et 1988,
Mart\'\i n \et 1992a, Pallavicini \et 1992, Bouvier \& Appenzeller 1992).
Low-mass stars spend their first few million years of life as T~Tauri stars
(TTS), and gradually evolve into post T~Tauri stars (PTTS), which represent a
longer, yet scarcely observed (Jones \& Herbig 1979), phase of evolution.

T~Tauri stars are commonly divided in two subtypes: classical TTS (CTTS), which
generally accommodate to the original criteria defining the T~Tauri class
(Herbig 1962), and weak TTS (WTTS), which lack most of the properties typical
of the CTTS. However, the distinction between CTTS and WTTS is sometimes
complicated because the emission lines span a continuous range of strengths and
are usually variable. Furthermore,  there is a region in the
Hertzsprung-Russell (H-R) diagram where CTTS and WTTS are mixed,  which
corresponds to totally convective PMS Hayashi tracks. In the region of the H-R
diagram closer to the main-sequence, on the radiative PMS Henyey tracks, only
WTTS and PTTS are found (e.g. Mart\'\i n \et 1992a).

Theoretical evolutionary models face considerable uncertainties to theorize the
evolution of PMS stars (cf. Mazzitelli 1989). Observations of lithium
abundances in TTS are a test to PMS models because they can be confronted with
Li depletion predictions. The problem of Li burning has been considered in a
number of theoretical papers, from the pioneering work of Bodenheimer 1965, to
recent papers, e.g., Pinsonneault \et 1990, Swenson \et 1990 and D'Antona \&
Mazzitelli 1993. The importance of Li depletion in PMS stars is not only
limited to PMS evolution itself, but it has far-reaching implications. For
example,  in the calibration of Li depletion mechanisms on the main-sequence
(gravity waves, diffusion, winds, etc), the evolution of lithium in the Galaxy
(the initial Li content of TTS is a measurement of the current Li abundance in
molecular clouds), and  the  identification of substellar objects (preservation
of Li in the brown dwarf regime, Rebolo \et 1992, Magazz\`u \et 1993a).

{}From an observational point of view, the high resolution spectroscopic data
needed for performing lithium abundance studies in large samples of TTS have
only recently become available (Magazz\`u \& Rebolo 1989 and Strom \et 1989a).
Strom \et made a comparative study of \li\ equivalent width measurements in TTS
and $\alpha$Per cluster members (age $\approx$ 50 Myr). Their main results
were: (1) The maximum Li abundances found in TTS were at least 0.3 dex. higher
than the maximum abundances in $\alpha$Per stars, implying that either (a)
there is significant PMS Li depletion in stars more massive than the Sun, or
(b) the molecular clouds associated to TTS are lithium rich with respect to
$\alpha$Per. (2) For TTS less massive than the Sun the scatter in Li abundances
was about 1 dex., but no correlation of Li with age and mass was found. In a
more recent work Basri \et 1991 showed that the Li abundances of Strom \et 1989
had to be taken with extreme caution because of the high uncertainties involved
in the analysis. The main sources of uncertainty were veiling corrections,
effective temperatures and the details of the abundance analysis procedure.

Very recently Magazz\`u \et 1992  improved the analysis of Li abundances in
TTS, and reached the following conclusions: (1) The initial Li abundance of TTS
is log N(Li)=3.2\x0.2 (in the usual scale of
log~N(Li)~=~12~+~log(N$_{Li}$/N$_H$), in agreement with the cosmic Li abundance
(interstellar medium, young clusters, meteorites). (2) There are hints of
correlations between Li abundance, age and mass, but, unfortunately very few
stars in their sample showed Li depletion, preventing detailed comparison with
model predictions.

The latest published work on lithium in TTS is that of King 1993. His PMS stars
belong to Orion Ic and have higher masses than previous works
(1.5-2.8~M$_\odot$). King obtains Li abundances in the range  4.1$\ge$log
N(Li)$\ge$1.2, and interprets these results in terms of PMS Li depletion, with
initial abundance of order log~N(Li)=4. However, King's abundances have been
estimated from curves of growth published by Strom \et 1989a, which have been
shown to be inadequate by the considerations in Duncan 1991 and Magazz\`u \et
1992. Hence, the results of King 1993 are probably affected by systematic
effects.

The value of the initial lithium abundance of T Tauri stars remains
controversial, and the details of when lithium depletion starts and how it
proceeds with PMS evolution are still to be addressed by the observations. We
have tried to approach these problems by concentrating our observations on
WTTS, because of a number of reasons: (1) They have not been studied in detail
by previous works, which mainly dealed with CTTS. (2) They are free from
optical veiling effects which are a source of uncertainty in CTTS (Basri \&
Batalha 1990, Basri \et 1991). (3) Their spectral types are more reliable than
those of CTTS. (4) Their atmospheres of are in a state near to equilibrium
(Finkenzeller \& Basri 1987). (5) Their positions in the H-R diagram are little
affected by uncertain corrections of disk luminosity. (6) They occupy a a wider
space in the H-R diagram than CTTS do. Therefore, the study of WTTS can be more
informative and reliable, and possibly a necessary step towards understanding
the complexity of the CTTS.

The Li abundances in WTTS presented in this paper provide clues about the
initial conditions of PMS Li burning. In the next paper of this series
(Garc\'\i a L\'opez \et 1993, Paper II) we will explore the final conditions of
PMS Li burning through the study of Li abundances in low-mass members of the
Pleiades cluster.

\section{Observations}
\label{sec2}

\subsection{Program stars}

Our sample has been selected from the Herbig \& Bell 1988 catalogue (hereafter
HBC). Usually  WTTS are separated from CTTS attending to the observational
criterion of \ha~equivalent width less than 10~\AA\ \footnote {10~\AA\ = 1~nm,
10~m\AA\ = 1~pm. Following the recommendations of the Executive Committee of
the IAU, we adopt the units of the International System.}. The lack of a full
physical understanding of the differences between the two subclasses of T~Tauri
stars prevents the definition of a clear borderline between them. Hence we
chose to observed mainly TTS with small \ha~equivalent width because we
intended to focus on WTTS which had no lithium measurements at high resolution.
Our limiting red magnitude was set at R$\approx$14, and we only took stars with
spectral type later than G0 and declinations higher that -20\deg. Thus, the
bulk of the WTTS observed by us are in the Taurus star-forming region. There
are no other selection effect in our sample apart from those exposed above and
those inherited from the surveys where the stars were discovered (for instance
bias towards strong X-ray emitters).

For the photometric observations we chose WTTS in the HBC that lacked a
complete set of UBVRI colors. Our aim was to derive the bolometric luminosities
from our own photometric data with the ultimate goal of placing the WTTS in the
HR diagram. Most of the stars observed photometrically were also observed
spectroscopically.

\subsection{Spectroscopy}

Spectroscopic observations were conducted during three runs at the 2.5~m Isaac
Newton telescope (INT), two runs at the 4.2~m William Herschel telescope (WHT),
and one run at the 2.5~m Nordic Optical Telescope (NOT). A summary of the
observing log is presented in Table~1. The name of the star in column~1 is that
of the first name given in the HBC.  Two stars are grouped together if they
were observed simultaneously along the slit ,i.e. they have angular separations
between 2 and 120\sec.  The time (in seconds) in column~3 is the total exposure
time on the object, i.e. summing up the times of succesive exposures. In
column~4 we listed the telescope where the observation was made. The instrument
used at the INT was the Intermediate Dispersion Spectrograph, while at the WHT
we used ISIS (Unger \et 1988), and at the NOT we employed the echelle
spectrograph IACUB (McKeith \et 1993). The various nominal dispersions and
wavelength coverages in column~5 and 6 respectively, result from  different
combinations of gratings, cameras and CCD detectors. The FWHM effective
resolutions are in the range 0.02 to 0.075~nm. When the spectral range was
21.2~nm or greater we could observe simultaneously \ha\ and the \li\ region
(see Figures 1 to 3), but when it was smaller we missed \ha\ (see Figure~4).
The last column of Table~1 provides and estimate of the final signal to noise
ratio achieved on each object. The S/N is probably underestimated for the
M-type stars because the true continuum is depressed by molecular absorption.

The data reduction was made with standard procedures available from the IMRED
and the TWODSPEC packages in IRAF\footnote {IRAF is distributed by National
Optical Astronomy Observatory, which is operated by the Association of
Universities for Research in Astronomy, Inc., under contract with the National
Science Foundation.}. Each image was de-biased, flat fielded, and background
subtracted. The spectra were wavelength calibrated using exposures of a CuAr
lamp taken after each object frame. The rms of the third order polynomial
dispersion solutions were lower than 0.001~nm.

\begin {table*} \caption[]{Spectroscopic Observations}
\label{tab1}
\begin{flushleft} \begin {tabular}{lllcccc}
& & & & & & \\
\hline Name & Date & t$_{exp}$  & Tel. & Disp. & Range & S/N \\
HBC &  & (s) &  & (nm pix$^{-1}$) & (nm) & continuum \\
\hline V819~Tau & 08.03.90 & 1800 & INT & 0.022 & 13.0 & 50  \\
CZ,DD Tau  & 09.03.90 & 1800 & INT & 0.036 & 21.2 & 50,80  \\
040234+2143 & 10.03.90 & 1800 & INT & 0.022 & 13.0 & 50  \\
040047+2603W,E & 27.12.90 & 3600 & WHT & 0.037 & 43.7 & 30,35  \\
041559+1716 & 27.12.90 & 2400 & WHT & 0.037 & 43.7 & 105  \\
V927 Tau & 28.12.90 & 3600 & WHT & 0.037 & 43.7 & 35  \\
Anon 1 & 28.12.90 & 3000 & WHT & 0.037 & 43.7 & 75  \\
V710 Tau A,B & 29.12.90 & 1600 & WHT & 0.037 & 43.7 & 40,35  \\
IS Tau & 30.12.90 & 1800 & WHT & 0.037 & 43.7 & 60  \\
1E0255+2018 & 30.12.90 & 900 & WHT & 0.037 & 43.7 & 95 \\
            & 31.12.90 & 1000 & INT & 0.036 & 21.2 & 50 \\
V807,GH Tau & 31.12.90 & 1200 & INT & 0.036 & 21.2 & 130,50  \\
V928 Tau & 31.12.90 & 2800 & INT & 0.036 & 21.2 & 70  \\
Wa Tau/1 & 31.12.90 & 1200 & INT & 0.036 & 21.2 & 200  \\
035120+3154SW,NE & 31.12.90 & 1200 & INT & 0.022 & 13.0 & 80,80  \\
042916+1751 & 31.12.90 & 1200 & INT & 0.036 & 21.2 & 70  \\
V773 Tau$^*$ & 02.01.91 & 1200 & INT & 0.022 & 13.0 & 150  \\
Hubble 4 & 03.01.91 & 1200 & INT & 0.022 & 13.0 & 80  \\
HV Tau & 05.01.91 & 3600 & INT & 0.022 & 13.0 & 50  \\
035135+2528NW,SE & 23.01.92 & 1500 & INT & 0.037 & 46.8 & 80,100 \\
040142+2150SW,NE & 23.01.92 & 1800 & INT & 0.037 & 46.8 & 20,20 \\
040012+2545N,S & 24.01.92 & 1200 & INT & 0.037 & 46.8 & 50,50 \\
042835+1700 & 24.01.92 & 1500 & INT & 0.037 & 46.8 & 80 \\
            & 28.02.93 & 1000 & WHT & 0.037 & 47.3 & 100 \\
045251+3016 & 24.01.92 & 1400 & INT & 0.037 & 46.8 & 110 \\
LkCa21 & 24.01.92 & 1800 & INT & 0.037 & 46.8 & 90 \\
032641+2420 & 17.12.92 & 3600 & NOT & 0.014 & 4.8 & 35 \\
BD+24.676 & 17.12.92 & 1800 & NOT & 0.014 & 4.8 & 90 \\
041529+1652 & 28.02.93 & 1000 & WHT & 0.037 & 47.3 & 100 \\
043124+1824 & 28.02.93 & 1000 & WHT & 0.037 & 47.3 & 100 \\
\hline
\end{tabular}

\medskip

Note: ($^*$) V773~Tau was re-observed five times with same exposure time
and resolution on the nights from 3 to 5 January 1992.

\end{flushleft}

\end{table*}

In Fig.~1 we display the spectra of program stars observed at the WHT. Note
that the position of the \li\ line is marked. We have ordered the stars in this
and the following figures from earlier to later spectral type (downwards). Note
the absence of the \lii\ resonance line in the spectra of the two components of
the visual binary NTTS040047+2603. In Fig.~2 and Fig.~3 we display in a similar
fashion spectra obtained at the INT. Note that Wa~Tau/c and NTTS042835+1700
show double lines (see individual remarks at the end to this section). In
Fig.~4 we present our higher resolution spectra. Note that V773~Tau has
strongly asymmetric line profiles. Due to the scarcity of known double-lined
spectroscopic binaries among T~Tauri stars (Mathieu \et 1989), the finding that
NTTS042835+1700 and V773~Tau are likely to be such binaries is of special
interest and requires further observational effort to determine the orbital
parameters.

\subsection{Photometry}

Broad band UBVRI photometry in the Johnson-Cousins standard system
was obtained at the 1~m Jacobus Kapteyn telescope. We used the
People's photometer (cf. Mart\'\i n \et 1992a and references therein) for 13
stars and the CCD camera for 2 close
visual binary systems.
In Table~2 we have listed the Julian dates,
visual magnitudes and colors of the objects observed with the photoelectric
photometer, and
in Table~3 we give the results from the CCD photometry. The visual
binaries HV~Tau and V710~Tau have angular separations smaller than
10\sec (see next subsection for more comments). An account of our
CCD photometric observations of close visual pre-main sequence binaries
is given elsewhere (Mart\'\i n 1993a).

\begin {table*}
\caption[]{Photometric Observations}
\label{tab2}
\begin{flushleft} \begin {tabular}{lllcccc}
& & & & & & \\
\hline Object & J.D. & V & U-B & B-V & V-R & V-I \\
\hline 1E0255+2018 & 260.359 & 11.99 & 0.89 & 1.19 & 0.72 & 1.37 \\
                   & 260.364 & 11.99 & 0.88 & 1.21 & 0.72 & 1.38 \\
                   & 261.390 & 12.01 & 0.89 & 1.19 & 0.72 & 1.38 \\
                   & 262.395 & 12.02 & 0.87 & 1.16 & 0.70 & 1.35 \\
V773~Tau & 260.377 & 10.88 & 1.11 & 1.36 & 0.86 & 1.70 \\
         & 261.395 & 10.82 & 1.12 & 1.32 & 0.85 & 1.69  \\
         & 262.431 & 10.87 & 1.13 & 1.35 & 0.86 & 1.69  \\
         & 263.563 & 10.84 & 1.12 & 1.38 & 0.86 & 1.71  \\
V807~Tau & 260.385 & 11.22 & 0.27 & 1.23 & 0.82 & 1.69  \\
         & 263.440 & 11.27 & 0.48 & 1.25 & 0.82 & 1.69  \\
Hubble 4 & 260.402 & 12.74 & 1.16 & 1.59 & 1.06 & 2.18  \\
       & 263.433 & 12.75 & 1.38: & 1.55 & 1.07 & 2.16  \\
Anon 1 & 260.415 & 13.52 & 1.61 & 1.79 & 1.24 & 2.54  \\
       & 262.436 & 13.56 & 1.64: & 1.75 & 1.24 & 2.53  \\
LkCa14 & 260.425 & 11.68 & 1.01 & 1.23 & 0.72 & 1.39  \\
IS~Tau & 261.439 & 15.44: &  & 2.10:: & 1.46: & 2.87:  \\
V928~Tau & 261.452 & 14.02 &  & 1.79: & 1.21 & 2.50  \\
         & 263.453 & 14.07 & 1.52:: & 1.73: & 1.22 & 2.49  \\
HV~Tau & 262.448 & 14.43 & 1.05: & 1.69 & 1.31 &   \\
         & 263.446 & 14.38 & 1.10: & 1.65: & 1.30 & 2.81  \\
VY~Tau & 262.455 & 13.66 & 1.11: & 1.45 & 0.97 &   \\
St 34 & 262.464 & 14.70 &  & 1.23: & 1.04 &   \\
FF~Tau & 264.456 & 13.96 & 1.75:: & 1.82: & 1.18 & 2.35  \\
Wa~Tau/1 & 264.543 & 10.34 & 0.49 & 0.94 & 0.50 & 0.90  \\
\hline
\end{tabular}

\medskip

Notes: J.D. = Julian Date - 2448000.0 . The photometric system
is the standard Jonhson-Cousins UBVRI. Average errors are less
than 0.05 mag., except where marked with ``:" for errors between
0.05 and 0.1 mag. and ``::" for errors between 0.1 and 0.3 mag.

\end {flushleft}

\end{table*}

\begin {table} \caption[]{Photometric Observations (CCD) }
\label{tab3}
\begin{flushleft} \begin {tabular}{lcccc}
& & & & \\
\hline Object & J.D. & V & V-R & V-I \\
\hline DO~Tau & 272.430 & 13.84 & 1.38 & 2.54   \\
HV~Tau A,B & 272.430 & 14.49 & 1.45 & 3.01   \\
HV~Tau C & 272.430 & 17.25:: & 1.44:: & 2.32::   \\
V710~Tau A & 336.313 & 14.33 & 1.18 &   \\
V710~Tau B & 336.313 & 14.48 & 1.21 &   \\
\hline
\end{tabular}

\medskip

Notes: J.D. = Julian Date - 2448000.0 . The photometric system
is the standard Jonhson-Cousins UBVRI. Average errors are less
than 0.05 mag., except where marked with ``::"
for errors between 0.1 and 0.3 mag.

\end {flushleft}

\end{table}

\subsection{Remarks on individual stars}

A by-product of our observations is varied information on stars
with different peculiarities.

1. {\bf CZ~Tau and DD~Tau}: This couple of TTS (separation $\sim$ 1\min) have
very different spectral properties. CZ Tau is a WTTS while DD Tau is a CTTS.
G\'omez de Castro \& Pudritz 1992 have drawn attention on DD~Tau because it may
be the first case where the forbidden line emission region of a T~Tauri wind
has been resolved into focal nodes. We find that the radial velocity (in the
local standard of rest) obtained from the \lii\ resonance line in the spectrum
of DD~Tau is 27\x5~\kms, while that of CZ~Tau is 11\x6~\kms. The radial
velocity of the molecular cloud around DD~Tau is 6.7~\kms (Edwards \& Snell
1982). Hence, the radial velocity of DD~Tau shows a significant discrepancy
with respect to the cloud. This could be an indication that DD~Tau is a
spectroscopic binary, but should be confirmed by monitoring of radial velocity
variations.   We also note that in DD~Tau the [SII] emission around 673.0~nm
(see~Fig.~2) is not symmetrical; the blue peak is about 2 times higher than the
red one.

2. {\bf V710~Tau~A and B}: Cohen \& Kuhi 1979 estimate a difference in V of
about 1 magnitude between the 2 components of this close visual binary
(separation about 3\sec). We find a much smaller difference (see Tab.~3), which
could be an indication that one or both components are highly variable. Our
observations (both spectroscopic and imaging) were made under good seeing
conditions. Blending of the two components was negligible.

3. {\bf 1E0255+2018}: This strong x-ray emitter (Fleming \et 1989) has been
reported to be a close visual binary with nearly-identical components (Mart\'\i
n \et 1992b). The spectrum in Fig.~2 is the combination of the 2 components
because we could not resolve them on the slit.

4. {\bf Wa~Tau/1}: Our spectrum reveals that this is a double-lined
spectroscopic binary (Fig.~2). We do not detect the \lii\ line in any of the 2
components.  The weak \ha\ emission may be due to chromospheric activity in a
close binary system. In order to test this hypothesis Mathieu 1992 has
re-observed this star and reports a short orbital period and center-of-mass
velocity inconsistent with membership to the Taurus clouds. Thus, we exclude
this star from our analysis.

5. {\bf V773~Tau}: Three of our spectra show asymmetric line-shapes very
suggestive of a double-lined pattern almost resolved. In Fig.~4 we present one
of the spectra where this is clearly seen. Because of the uncertainty
introduced by the different strengths of the photospheric lines of each
component, we will not consider this star further in our analysis of Li
abundances.

6. {\bf HV~Tau}: It is a triple system composed of two close stars (resolved
using speckle techniques) and a fainter visual companion (Simon \et1992). We
have used our CCD data to derive the luminosity of HV~Tau~AB. Spectroscopic
observations  of the faint visual companion (HV~Tau~C) are under analysis
(Magazz\`u \et1993b).

7. {\bf NTTS42835+1700}: New double-lined spectroscopic binary (Fig.~3). The
photospheric lines of both components have similar depths. We include this star
in our analysis under the assumption that the components are identical. Mathieu
1992 reports that NTTS42835+1700 does not show double lines in 9 spectra he has
inspected. In our spectrum of February 1993 (see Table~1) the star is also
single lined. This suggest that the binary orbit may have a high eccentricity.

8. {\bf BD+24.676}: The PMS evolutionary status of this star has been contested
by Mart\'\i n 1993b, who shows that it is in fact a spectroscopic binary with
one evolved component. Thus, we exclude it from the abundance analysis.

\section{Analysis}
\label{sec3}

\subsection{Equivalent width measurements}

In Table~4 we present equivalent widths of \ha~(usually in emission), the \lii\
resonance line and some neighbouring strong absorption lines in the spectra of
all program stars except the two DLS binaries V773~Tau and Wa~Tau/1 (see
remarks in previous section).  The equivalent widths of NTTS42835+1700 were
measured on the single-lined spectrum.

\begin {table*} \caption[]{Equivalent widths of program stars (pm)}
\label{tab4}
\begin{flushleft} \begin {tabular}{lccccc}
& & & & & \\
\hline Object & H$_\alpha$ & NiI$\lambda$664.36
& FeI$\lambda$667.80 & \lii$\lambda$670.78 & CaI$\lambda$671.77  \\
HBC & (nm) & (pm) & (pm) & (pm) & (pm) \\
\hline Anon 1 & -0.135 & 12.1 & 12.6 & 47.8 & 26:  \\
CZ Tau  & -0.64 &  &  & 46: & 28:  \\
DD Tau  & -8.95 &  &  & 22.9 &   \\
GH Tau & -1.085 & 12: & 11: & 65.6 & 44:  \\
Hubble 4 &  &  & (-25.5) & 61.2 & 30.4  \\
HV Tau & -0.43 &  &  & 64: &  \\
IS Tau & -1.305 & 17.1 & (1:) & 61.6 & 26.4  \\
Lk Ca 21 & -0.55 & 9.4 & 25: & 71.5 &   \\
V710 Tau A & -3.87 & 11.2 &  & 55.4 & 31:  \\
V710 Tau B & -0.36 & 13.1 &  & 58.2 &   \\
V807 Tau & -1.58 & 10.4 & (7.2) & 48.8 & 23.7  \\
V819 Tau &  &  & 27.0 & 62.2 & 26.5  \\
V927 Tau & -0.87 & 6: & & 51: &   \\
V928 Tau & -0.083 & 14.1 & (-16.2) & 63.9 & 44:  \\
1E0255+2018 & -0.52 & 17.9 & 27.3 & 47.5 & 25.0 \\
032641+2420   &  &  &  28.3  & 35.2 & 25: \\
035120+3154SW &  &  & 17: & 13: & 18: \\
035120+3154NE &  &  & 16.2 & 22.3 & 16.5 \\
035135+2528NW & 0.47: & 15: & 33.7 & 24.0 & 28: \\
035135+2528SE & 0.18: & 12.3 & 29.5 & 28.3 & 24.5 \\
040012+2545N+S & -0.030 &  & 32: & 36: & 34: \\
040047+2603W & -0.37 & 12.8 &  & \lee7 &  \\
040047+2603E & -0.85 &  9.5 &  & \lee6 &  \\
040142+2150SW & -1.25 &  &  & \lee25 &  \\
040142+2150NE & -0.66 &  &  & \lee30 &  \\
040234+2143 &  &  & (-8.7) & 34.3 &  \\
041559+1716 & -0.14 & 18.1 & 28.6 & 46: & 27:  \\
041529+1652 &  0.05 & 16.0 & 29.4 & 18.6 & 24.6  \\
042835+1700 & -0.037 & 13.0 & 26.5 & 11.0 & 28.8  \\
042916+1751 & -0.056 & 22.1 & 31.6 & 49.4 & 28.5  \\
043124+1824 &  0.09: & 13.2 & 19.5 & 23.0 & 21.0  \\
045251+3016 & -0.055 & 16.4 & 26.8 & 52.3 & 29.3  \\
\hline
\end{tabular}

\medskip

Notes: Negative equivalent widths indicate line emission.
Numbers in parenthesis mean  that the
FeI$\lambda$667.80 line is filled or disappears due to
HeI$\lambda$667.81 emission. Measurement errors are less
than 5\%, except where marked with a colon (error about 10\%).

\end {flushleft}

\end{table*}

We estimate an average error bar of 5\% in the equivalent width values at
1$\sigma$, coming mainly from uncertainty in the continuum placement. The
metallic lines could not be accurately measured for spectral types later than
about M1 because of the continuum modulation due to deep molecular bands and
because they get weaker as the excitation temperature decreases.

We have compared our \lii\ measurements with those of different authors and
find an agreement better than 15\% for one star in common with Hartmann \et1987
and six in common with Walter \et 1988. However, for other  eight stars in
common with the latter authors we find larger discrepancies, mainly in the
sense that we obtain smaller equivalent widths.  Walter \et1988 caution that
their measurements are very uncertain, and thus we believe that there is no
inconsistency with our results.

Those stars in our sample with \ha~equivalent width stronger than -1.50~nm
(-15~\AA), namely DD~Tau, V710~Tau~A and  V807~Tau, are not considered further
because we want to limit the study to moderate emission TTS. The stars GH~tau,
IS~Tau and NTTS040142+2150SW have \ha~strengths at the border of what may be
regarded acceptable for a WTTS. We accept them because their \ha~flux is
moderate and they do not show any other emission lines in our spectral range.

In Table~5 we show  equivalent widths of \ha~and \lii\ for WTTS that have not
been observed by us. The data are taken from published works  quoted at the
bottom of the table. For stars with more than one \lii\ measurement we place
equal weight on all the data and take the mean as the value to be used for
calculating the Li abundance. These stars from the literature represent 42\% of
the total sample of WTTS for which we derive Li abundances. We will refer to
our program stars plus those taken from other works as the extended sample.

\begin {table*} \caption[]{Equivalent widths from the literature (pm)}
\label{tab5}
\begin{flushleft} \begin {tabular}{lccc}
\hline Object & H$_\alpha$ & \lii$\lambda$670.78 & Source \\
HBC & (nm) & (pm) & \\
\hline DI Tau & -0.10 & 69 & S \\
IP Tau & -1.16,-0.80 & 48,52 & HSS,S \\
IW Tau & -0.37 & 44 & HSS \\
Lk Ca 1 & -0.28 & 56 & HSS \\
Lk Ca 3 & -0.14,-0.40 & 55,57 & HSS,S \\
Lk Ca 4 & -0.49,-0.50 & 71,51 & HSS,S \\
Lk Ca 5 & -0.25 & 55 & HSS \\
Lk Ca 7 & -0.53,-0.30,-0.37 & 63,52,55,60,59 & BMB,G,HSS,S,W \\
Lk Ca 14 & -0.09 & 60 & HSS \\
Lk Ca 15 & -1.27 & 47 & HSS \\
Lk Ca 19 & -0.05 & 45,48,43 & BMB,G,HSS \\
S-R 12 & -0.36 & 70 & MRP \\
Sz 65 & -0.33 & 61 & FB \\
Sz 68 & -0.68 & 42 & FB \\
Sz 82 & -0.66 & 57 & FB \\
UX Tau A & -0.665 & 43 & MMR \\
UX Tau B & -0.35 & 60 & MMR \\
V827 Tau & -0.30 & 57 & S \\
V830 Tau & -0.20 & 64,65 & BMB,S \\
V836 Tau & -0.70 & 57 & S \\
034903+2431 & -0.16 & 31,35 & BMB,W \\
042417+1744 & -0.05 & 28,28 & BMB,W \\
043230+1746 & -0.90 & 61,57 & G,W \\
\hline
\end{tabular}

\medskip

Key to references: BMB = Basri, Mart\'\i n \& Bertout 1991,
FB = Finkenzeller \& Basri 1987,
G = G\'omez et al. 1992, HSS = Hartmann, Soderblom \& Stauffer 1987,
MMR = Magazz\`u, Mart\'\i n \& Rebolo 1991,
MRP = Magazz\`u, Rebolo \& Pavlenko 1992,
S = Strom et al. 1989a,
W = Walter et al. 1988.

\medskip

Notes: Equivalent widths are given in pm and wavelengths in nm.
A minus sign in the equivalent width means that the line is in
emission.

\end {flushleft}
\end{table*}

\subsection{Effective temperatures}

T~Tauri stars are in general more luminous than main-sequence stars of the same
spectral type. Hence, TTS cannot be regarded as luminosity class~V stars. In
fact, no definite class can be assigned to them because they span a wide range
of luminosities. Relationships between the semi-quantitative parameters
luminosity class and spectral type with the quantity effective temperature are
needed for placing TTS in the H-R diagram, comparing with theoretical tracks
and isochrones and deriving surface chemical abundances. The relationship
adopted by  Cohen \& Kuhi 1979, which is basically the same of Johnson 1966 for
class~V stars,  is still widely used in PMS studies, but more recent
determinations are now available (Bessell 1979, de~Jager \& Nieuwenhuijzen
1987, Bessell 1991). Magazz\`u \et 1991 and Magazz\`u \et 1992 have adopted the
relationship of spectral type with temperature of  de~Jager \& Nieuwenhuijzen
1987 for luminosity class~IV. This luminosity class may be roughly adequate for
TTS lying in the upper part of PMS evolutionary tracks, but not for any
T~Tauri star. We prefer not to assume a priori any luminosity class, but
instead to treat each TTS separately. We seek to obtain an effective
temperature from the spectral type and the luminosity. For example, what is the
T$_{\rm eff}$ of a K7 T~Tauri with log~(L/L$_\odot$)=0~?. de~Jager \&
Nieuwenhuijzen 1987 have defined the parameters s and b to quantify the
spectral type and luminosity class respectively, and with the aid of their
Table~6 it is possible to convert the spectral type and luminosity to values of
s and b. Using such values, the  T$_{\rm eff}$ follows directly from   Table~5
and Fig.~5 of de~Jager \& Nieuwenhuijzen.

In our Table~6 we present various spectral type to T$_{\rm eff}$ relationships.
It is interesting to compare T$_{\rm eff}$  values obtained from the scale
adopted by Cohen \& Kuhi with those obtained from other scales. For instance,
if we compare with de~Jager \& Nieuwenhuijzen class V, we obtain the following
differences: for spectral types G0-G2, K4-K5 and M1-M3 less than 50~K, but for
spectral types in the range G4-K3 de Jager \& Nieuwenhuijzen T$_{\rm eff}$
values are always smaller than those of Cohen \& Kuhi up to 136~K (at G8),
whereas  for K7 and M2-M5 holds the reverse situation, with differences up to
150~K (at K7).

\begin {table}
\caption[]{Spectral Type - T$_{\rm eff}$ scales}
\label{tab6}
\begin{flushleft} \begin {tabular}{lcccc}
\hline Sp.T. & C.K. & B. & de J.N. IV & de J.N. V  \\
\hline
G0 & 5902 & 6000 & 5636 & 5943  \\
G2 & 5768 &      & 5458 & 5794  \\
G4 &      &      & 5284 & 5636  \\
G6 &      & 5500 & 5114 & 5475  \\
G8 & 5445 &      & 4943 & 5309  \\
K0 & 5236 &      & 4775 & 5150  \\
K1 & 5105 &      & 4624 & 4989  \\
K2 & 4954 & 5000 & 4480 & 4833  \\
K3 & 4775 &      & 4335 & 4690  \\
K4 & 4581 & 4500 & 4207 & 4540  \\
K5 & 4395 &      & 4080 & 4405  \\
K7 & 3999 & 4000 & 3870 & 4150  \\
M0 & 3917 & 3800 & 3630 & 3837  \\
M1 & 3681 & 3650 & 3507 & 3664  \\
M2 & 3499 & 3500 & 3411 & 3524  \\
M3 & 3357 & 3350 & 3341 & 3404  \\
M4 & 3228 & 3150 & 3280 & 3288  \\
M5 & 3119 & 3000 & 3221 & 3170  \\
\hline
\end{tabular}

\medskip

Key to references: C.K. = Cohen \& Kuhi 1979, B. = Bessell 1979
(for G/K spectral types) and Bessell 1991 (for M spectral types),
de~J.N. = de Jager \& Nieuwenhuijzen 1987.

\end{flushleft}

\end{table}

\subsection{Luminosities and positions in the H-R diagram}

Spectral types and bolometric luminosities for the majority of the WTTS have
been taken from the literature. The references are given at the end of Table~7.
We have classified IS~Tau and 1E0255+2018 as a K7 and a K6 star respectively
through comparison of their line depths in the spectral region 666-675~nm with
that of other WTTS of known spectral type.  The star LkCa21 has been classified
as M3 on the basis of the photometry reported in Bouvier \et 1993b. Using our
photometry (Tables~2 and 3) and that of Bouvier \et 1993b (only for LkCa21), we
have derived luminosities for Anon~1, Hubble~4, HV~Tau, LkCa14, LkCa21,
V710~Tau~B and 1E0255+2018 in the following way. A red color excess was derived
via comparison of the observed (V-R) color with that expected from the
spectraly type (Bessell 1979, 1991). (V-R) was chosen to minimize the
contribution of possible blue and near-IR excesses (Strom \et 1989b). The
absolute extinction was obtained from the calibration of color excess vs.
standard interstellar extinction given by Rieke \& Lefobski 1985. The
unreddened I color together with the bolometric correction corresponding to the
spectral type (Bessell 1991), and a distance of 150~pc to the Taurus clouds,
were used for obtaining the absolute luminosity. The visual binary 1E0255+2018
was assumed to be also at 150~pc, and the luminosity was equally divided
between both components. We did the same with the double-lined spectroscopic
binary NTTS~042835+1700. Our method of deriving luminosities gives consistent
results when compared with those employed by Strom \et 1989b and Walter \et
1988. Very recently, Simon \et 1993 have corrected the luminosity of TTS from
the contribution of IR companions. We have adopted their luminosities for the
stars in common.

The spectral types and luminosities adopted for the extended sample of WTTS are
given in tables~7 and 8. Effective temperatures were calculated as explained in
the previous section. Masses and ages were inferred from the position of the
stars in the H-R diagram and comparison with  PMS models. We have taken those
of  D'Antona \& Mazzitelli 1993 (hereafter DAM) with updated opacities and
convection theory (DAM's set number 1). We also compared the H-R locations of
the WTTS with the non-rotating models of Pinsonneault \et 1990 (hereafter PKD)
and found only modest differences in masses (less than 0.1~M$_\odot$) and ages
(up to a factor~2). In Fig.~5 we plot the WTTS in the the H-R diagram, together
with the tracks and isochrones of DAM used to derive masses and ages.

\subsection{Calculation of lithium abundances}

Lithium abundances were obtained from comparison of the measured \li\
equivalent widths with curve of growth calculations in LTE and NLTE conditions.
We used the most recently available model atmospheres generated in LTE by
Kurucz 1992 with solar metallicity, and gravities and effective temperatures in
the range log~g=3.0-4.5 and T$_{\rm eff}$=3500-6000~K. The formation of the
\lii\ resonance line was followed to the uppermost atmospheric layers, up to
levels where the optical depth in the center of the \lii\ strongest component
of the resonance doublet was less than 0.05. For very high Li abundances it was
necessary sometimes to extrapolate to higher levels than considered in the
model atmospheres. The extrapolation was made in the same way as explained in
Magazz\`u \et 1992.

In Fig.~6 we show a set of the LTE and NLTE curves of growth employed in this
work. They were computed assuming  a microturbulence velocity of 2~\kms. The
computational method was described in Magazz\`u \et 1992 (and references
therein). However, the NLTE curves reported here  are more accurate than those
in Magazz\`u \et 1992 because they were made using a 20-level Li atom model,
rather than the simplified 6-level atom. We included 70 radiative and all
possible collision (with hydrogen and free electrons) transitions. All the
relevant references of cross sections, damping constants and oscillator
strengths are the same as in Magazz\`u \et 1992.

{}From Fig.~6 it can be seen that the NLTE corrections to the LTE calculations
are typically smaller than 0.1~dex. in log~N(Li). The \lii\ resonance line in
TTS is more sensitive to typical errors in  T$_{\rm eff}$ or even log~g than to
NLTE effects. An uncertainty of $\pm$~250~K in T$_{\rm eff}$, or 0.5 dex. in
log~g, translate into error bars in log~N(Li) of about 0.4 dex. and 0.1 dex.
respectively. The spectral types of WTTS are usually determined with accuracy
better than one spectral subclass, but as we saw before there can also be
errors inherent to the conversion from spectral type to T$_{\rm eff}$. We
assume an uncertainty of $\pm$~150~K in T$_{\rm eff}$, $\pm$~0.5 dex. in log~g,
and 10\% in observed equivalent width, which leads to a combined uncertainty in
log~N(Li) of $\pm$0.35~dex. at 1~$\sigma$.

\begin{table*}
\caption[]{Stellar parameters and lithium abundances of weak T Tauri stars}
\label{tab7}
\begin{flushleft} \begin {tabular}{lccccccccc}
\hline Name & Sp.T. & L/L$_\odot$ & Ref. &  T$_{\rm eff}$  & M/M$_\odot$  &
Age & log g & log N(Li) & log N(Li)  \\
& & & & (K) & & (Myr) & & LTE & NLTE \\
\hline Anon 1 & M0 & 0.77 & 1,2 & 3830 & 0.4 & 2 & 3.45 & 2.6 & 2.6 \\
CZ Tau  & M1.5 & 0.24 & 1,5 & 3590 & 0.35  & 2 & 3.78 & 2.05 & 2.1 \\
DI Tau & M0 & 0.73 & 1,5 & 3826 & 0.4  & 2 & 3.34 & 3.25 & 3.2 \\
GH Tau & M2 & 0.89 & 1,5 & 3520 & 0.25  & 0.4 & 3.03 & 2.9 & 2.9 \\
Hubble 4 & K7 & 0.95 & 1,2 & 4133 & 0.55 & 0.8 & 3.62 & 3.3 & 3.25 \\
HV Tau AB & M2 & 0.65 & 3,2 & 3521 & 0.25 & 0.5 & 3.17 & 2.8 & 2.8 \\
IP Tau & M0 & 0.5 & 1,1 & 3832 & 0.4 & 1 & 3.64 & 2.6 & 2.6 \\
IS Tau & K7 & 1.08 & 2,5 & 4130 & 0.5 & 0.5 & 3.50 & 3.2 & 3.15 \\
IW Tau & K7 & 1.12 & 1,4 & 4130 & 0.5 & 0.5 & 3.51 & 2.9 & 2.85 \\
Lk Ca 1 & M4 & 0.66 & 1,4 & 3288 & 0.2 & 0.5 & 2.95 &  &  \\
Lk Ca 3 & M1 & 0.98 & 1,5 & 3657 & 0.3 & 0.5 & 3.14 & 3.05 & 3.0 \\
Lk Ca 4 & K7 & 0.89 & 1,5 & 4130 & 0.55 & 1.0 & 3.65 & 3.3 & 3.25 \\
Lk Ca 5 & M2 & 0.38 & 1,4 & 3522 & 0.3 & 1 & 3.48 & 2.45 & 2.4 \\
Lk Ca 7 & K7 & 0.6 & 1,5 & 4133 & 0.6 & 2 & 3.86 & 3.25 & 3.2 \\
Lk Ca 14 & K7 & 0.88 & 2,2 & 4134 & 0.6 & 1 & 3.70 & 3.3 & 3.25 \\
Lk Ca 15 & K5 & 0.72 & 1,4 & 4385 & 0.8 & 3 & 4.01 & 3.1 & 3.0 \\
Lk Ca 19 & K5 & 1.55 & 4,4 & 4348 & 0.65 & 0.5 & 3.64 & 3.1 & 3.0 \\
Lk Ca 21 & M3 & 0.6 & 2,2 & 3400 & 0.2 & 0.5 & 3.05 & 2.9 & 2.9 \\
S-R 12 & M1 & 1.0 & 3,3 & 3660 & 0.3 & 0.5 & 3.13 & 3.15 & 3.1 \\
Sz 65 & K7/M0 & 0.35 & 5,5 & 3990 & 0.6 & 4 & 4.04 & 3.1 & 3.0 \\
Sz 68 & K2 & 3.0 & 5,5 & 4722 & 1.0 & 0.5 & 3.62 & 3.65 & 3.35 \\
Sz 82 & M0 & 0.5 & 5,5 & 3822 & 0.4 & 1 & 3.63 & 2.85 & 2.8 \\
UX Tau A & K2 & 1.3 & 1,1 & 4740 & 1.1 & 3 & 4.03 & 3.6 & 3.4 \\
UX Tau B & M1 & 0.5 & 1,1 & 3650 & 0.35 & 0.5 & 3.49 & 3.05 & 3.0 \\
V710 Tau B & M3 & 0.3 & 3,2 & 3402 & 0.2 & 1 & 3.35 & 2.4 & 2.4 \\
V819 Tau & K7 & 0.80 & 1,5 & 4136 & 0.6 & 1 & 3.66 & 3.3 & 3.25 \\
V827 Tau & K7 & 1.11 & 1,1 & 4130 & 0.5 & 0.5 & 3.52 & 3.25 &  3.2 \\
V830 Tau & K7 & 0.89 & 1,1 & 4134 & 0.6 & 1 & 3.69 & 3.5 & 3.35 \\
V836 Tau & K7 & 0.6 & 1,1 & 4139 & 0.6 & 2 & 3.87 & 3.2 & 3.15 \\
V927 Tau & M5.5 & 0.36 & 1,5 & 3101 & 0.15 & 0.5 & 2.99 &  &  \\
V928 Tau & M0.5 & 1.3 & 3,3 & 3745 & 0.3 & $<$0.1 & 3.06 & 3.1 & 3.05 \\
\hline
\end{tabular}

\medskip

References: 1. Strom \et 1989b, 2. This work, 3. Cohen \& Kuhi 1979,
4. Walter \et 1988, 5. Simon \et 1993 .

\end {flushleft}

\end{table*}

In Tables~7 and 8 we list the Li abundances of 46~WTTS and upper limits
for another 4~WTTS. Note that the Li abundances of
1E0255+2018, NTTS040012+2545 and NTTS042835+1700 refer to the two
nearly identical components of these binary systems and hence it
increases to 53 the total of WTTS considered here.
For the M3 stars ($\sim$3400~K) we extrapolated the predicted \li\
equivalent widths  according to a polynomial fit performed in the range
4000-3500~K, but no Li abundances
could be derived for LkCa1 and V927~Tau because their effective
temperatures are much lower than the model atmospheres available.

\begin{table*}
\caption[]{Stellar parameters and Li abundances of X-ray discovered WTTS}
\label{tab8}
\begin{flushleft} \begin {tabular}{lccccccccc}
\hline Name & Sp. & L/L$_\odot$ & Ref. & T$_{\rm eff}$ & M/M$_\odot$  &
Age & log g & log N(Li) & log N(Li)  \\
&  & & & (K) & & (Myr) & & LTE & NLTE \\
\hline 32641+2420 & K1 & 0.50 & 1, 1 & 4985 &  1.0 & 15 & 4.49 & 3.4 & 3.15 \\
34903+2431 & K5 & 0.32 & 1, 2 & 4391 & 0.85 & 15 & 4.39 & 2.25 & 2.3 \\
35120+3154SW & G0 & 0.77 & 1, 1 & 5943 & 1.05 & 50 & 4.63 & 3.1 & 3.0 \\
35120+3154NE & G5 & 0.56 & 1, 1 & 5600 & 1.0 & 50 & 4.64 & 3.2 & 3.1 \\
35135+2528NW & K3 & 0.22 & 1, 1 & 4690 & 0.75 & 40 & 4.62 & 2.15 & 2.2 \\
35135+2528SE & K2 & 0.33 & 1, 1 & 4835 & 0.85 & 30 & 4.55 & 2.8 & 2.7 \\
40012+2545N,S & K2 & 0.16 & 1, 1 & 4839 & 0.75 & 50 & 4.81 & 3.35 & 3.15 \\
40047+2603W & M2 & 0.20 & 1, 2 & 3524 & 0.3 & 2 & 3.76 & $\le$-0.2 & $\le$-0.2
\\
40047+2603E & M2 & 0.20 & 1, 2 & 3524 & 0.3 & 2 & 3.76 & $\le$-0.2 & $\le$-0.2
\\
40142+2150SW & M3 & 0.17 & 1, 1 & 3404 & 0.2  & 2 & 3.60 & $\le$0.5 & $\le$0.5
\\
40142+2150NE & M3 & 0.16 & 1, 1 & 3404 & 0.2 & 2 & 3.62 & $\le$1.0 & $\le$1.0
\\
40234+2143 & M2 & 0.18 & 1, 1 & 3524 & 0.3   & 3 & 3.81 & 1.3 & 1.35 \\
41529+1652 & K5 & 0.17 & 1, 1 & 4400 & 0.75 & 30 & 4.62 & 1.45 & 1.55 \\
41559+1716 & K7 & 0.40 & 1, 2 & 4142 & 0.75 & 6 & 4.14 & 2.7 & 2.65 \\
42417+1744 & K1 & 0.89 & 1, 1 & 4913 & 1.2 & 8 & 4.29 & 2.8 & 2.75 \\
42835+1700AB & K5 & 0.20 & 1, 1 & 4400 & 0.75 & 30 & 4.55 & 1.05 & 1.15 \\
42916+1751 & K7 & 0.62 & 1, 1 & 4138 & 0.6 & 2 & 3.85 & 3.15 & 3.05 \\
43124+1824 & G8 & 0.47 & 1, 1 & 5445 & 0.95 & 50 & 4.65 & 3.1 & 3.0 \\
43230+1746 & M2 & 0.22 & 1, 2 & 3524 & 0.3 & 2 & 3.74 & 2.55 & 2.5 \\
45251+3016 & K7 & 1.0 & 1, 1 & 4130 & 0.6 & 1 & 3.63 & 3.15 & 3.05 \\
1E0255+2018 & K6 & 0.6 & 3,3 & 4264 & 0.8 & 3 & 4.04 & 3.0 & 2.9 \\
\hline
\end{tabular}

\medskip

References: 1. Walter \et 1988, 2. Simon \et 1993, 3. this work .

\end {flushleft}

\end{table*}

\section{Discussion}
\label{sec4}

\subsection{The statistical distribution of lithium abundances}

\subsubsection{Homegeneity of the initial lithium abundance}

The statistical distribution of Li abundances (in NLTE) of the extended sample
of WTTS is shown in Fig.~7. About 85$\%$ of the WTTS have log~N(Li) in the
narrow range 3.4-2.9 . The distribution peaks around log~N(Li)=3.1, and we
interpret its shape as due to random errors around the inital Li abundance,
with a long extension towards lower abundances, due to physical processes of
PMS Li depletion. Thus, our analysis shows that the initial Li abundance in TTS
coincides with the maximum Li abundances measured in various astrophysical
contexts (open clusters, local interstellar medium, meteorites). Our result is
also consistent with those of Magazz\`u \et 1992 for a different sample of TTS
(mainly CTTS). Taken together, all these results support the view that the
lithium abundance of the interstellar medium around the Sun has remained fairly
constant during the last $\sim$~5~Gyr.

In order to explore the homogeneity of the initial Li abundance in WTTS we
restrict the sample to those stars with luminosity greater than 0.9L$_\odot$.
No  evolutionary model predicts any Li depletion in PMS stars of any mass at
such high luminosities. In this restricted sample, we find log~N(Li)=3.19 with
a r.m.s of 0.07 in LTE, and log~N(Li)=3.11 with a r.m.s of 0.06 in NLTE. These
numbers can be compared with the average LTE Li abundance in F-type members of
the $\alpha$~Per and Pleiades clusters which is log~N(Li)=3.0$\pm$0.1
(Boesgaard \et 1988). However, Mart\'\i n \& Rebolo 1993  found evidence for a
small amount of Li depletion in the cooler stars of the Boesgaard \et sample.
Using the same Li abundance code as we have used in this work, Mart\'\i n \&
Rebolo derived a mean NLTE Li abundance of log~N(Li)=3.13 with a r.m.s of 0.03
for the hotter stars of Boesgaard \et\, for which the theoretical models
predict negligible Li depletion. Hence, the initial Li abundance of our sample
of WTTS (mainly located in the Taurus molecular clouds) is the same as that of
hot stars in  $\alpha$Per and the Pleiades. In the next sections we will refer
to log~N(Li)=3.1 as the initial Li abundance.

The reasons why some authors have reported very high Li abundances in TTS are
mainly two: (1) The use of atmospheric models that do not include all the
layers where the \lii\ doublet is formed. This causes the computation to be
cutoff before the line has fully formed and leads to overestimating the Li
abundance required to produce strong Li lines. (2) The NLTE corrections, which
push the Li abundances towards lower values when the line is saturated. Both
these systematic effects conspire to produce the very high Li abundances
presented by Magazz\`u \& Rebolo 1989, Strom \et 1989a, Basri \et 1991 and King
1993. Our analysis, and those of Duncan 1991 and Magazz\`u \et 1992, have used
computations that include all the atmospheric layers where the Li doublet is
formed, and we consistently find lower Li abundances on the average.

\subsubsection{The decline of lithium}

The tail of WTTS at low Li abundances seen in Fig.~7 is a strong evidence that
processes of nuclear Li burning and convective material transport have been
effective in this stars. During PMS evolution the dominant mixing mecanism is
thought to be deep convection, evolving from total convection along the Hayashi
tracks to gradually shallower convection along the radiative Henyey tracks. In
this general framework Li is expected to be correlated with luminosity: as the
PMS star contracts towards lower luminosities its central pressure and
temperature increases and Li is destroyed. Of special interest is the
determination of the luminosity at which the depletion starts to be seen on the
stellar surface because it provides direct information on the central
temperature.

In Fig.~8 we have plotted the observed Li abundances against luminosities for
the WTTS of our extended sample. We feel that comparison of Li abundances with
luminosities is more meaningful than with ages or masses, because luminosity is
a direct observable, while ages and masses are model-dependent. However,
because of the great sensitivity of Li to mass, we take into account the
theoretical masses by giving different symbols in Fig. 8 to stars in 4 bins of
masses. There is a clear trend towards lower Li abundance at lower luminosity.
All the WTTS with log~N(Li)\lt2.4  have luminosities below 0.5~L$_\odot$.
Hence, we may take this luminosity as a conservative estimate of where Li
depletions begins to be significant. The largest Li depletions are found among
the WTTS with lowest luminosities. Hence, in a qualitative sense the observed
pattern of Li abundances is correlated with luminosity as theoretically
expected, but there are a number of problems when detailed comparison with
models is made.

We will enumerate the most important traits seen in Fig.~8:

\begin {enumerate}

\item  WTTS in the mass range 0.2-0.4~M$_\odot$ present an abrupt decline of Li
abundances below 0.4~L$_\odot$, impliying that at this luminosity the internal
conditions for Li burning are attained. The models predict that the luminosity
at the start of Li burning is about 0.1~L$_\odot$, i.e. shifted by a factor 4
from the observed turning point. Since these stars are totally convective, we
infer from the observations higher central temperatures in the luminosity range
0.4-0.2~L$_\odot$ than previously thought. This may not be surprising because
of the large theoretical uncertainties at such low masses. More computations at
these very low mass stars may provide useful tests to the opacities used by the
models.

\item In the range 0.4-0.6~M$_\odot$ we only have stars at luminosities greater
than 0.3~L$_\odot$, and they all have Li abundances near to the initial value.
This is not inconsistent with the models. It would be necessary to observe
stars at lower luminosities to test the models in this range of masses.

\item WTTS of about 0.8~M$_\odot$ (open squares in Fig.~8) show clear cases of
Li depletion. At these masses we have stars along all the PMS evolutionary
track, including the radiative approach to the ZAMS (Fig.~5). The non-rotating
models of PKD  for 0.8~M$_\odot$ predict Li depletions consistent with the
lower evelope of Li abundances. We also find consistency with the DAM model
using  the convection theory of Canuto \& Mazzitelli (CM) and Kurucz opacities
(set 2). However, using the same opacities the DAM model with mixing-length
(MLT) convection theory overestimates the Li depletion. This may suggest that
the CM theory is better than the MLT, but the DAM model with MLT treatment and
Alexander opacities (set 3) is also consistent with the obervations. Hence,
there is no clear difference in Li depletion predictions attributable to CM or
MLT convections theories and both of them seems a priori capable to fit the
maxima of Li depletion.

\item For the higher bin of masses (1.0-1.2~M$_\odot$) there are WTTS of widely
different ages (1-50~Myr), but we do not observe significant differences in Li
abundances. We interpret this result as evidence that the PMS Li burning at
these masses is less than about factor 2. The PKD and DAM models that predict
Li depletions realistic for a 0.8~M$_\odot$  PMS star, also predict Li
depletion of about 0.6 dex. at 1.0~M$_\odot$, which is not supported by the
observations. Mart\'\i n \& Rebolo 1993 have studied the secondary of the
eclipsing binary system EK~Cep, which is a 1.12~M$_\odot$ star with age about
20~Myr. They find a lithium abundance of log~N(Li)=3.1 and compare it with Li
abundances of $\alpha$~Per and Pleiades stars. Taken this into account and our
present results, we conclude that there is virtually no PMS Li burning at
masses between 1.2-1.0~M$_\odot$. The solar Li abundance shows a Li depletion
of about a factor 100, which must be a consequence of mixing during MS
lifetime.

\end {enumerate}

Considering the whole range of masses, 0.2-1.2~M$_\odot$, we have not
found a model able to provide an acceptable fit to
the global pattern of Li abundances. At 1.0~M$_\odot$ the models should
consider modifications in the input physics that reduce PMS Li burning.
For instance, using MLT convection theory, this would require a
lower value of the mixing parameter (l/H$_p$). At the lower
mass end 0.4-0.2~M$_\odot$ the discrepancy between models and observations
is very serious. This casts doubts on the estimates of masses and ages based
on current evolutionary tracks and isochrones for M-type T Tauri stars.

\subsection{Is there a link between PMS Li depletion and rotation?}

In Fig.~9 we have plotted log N(Li) vs. v~sin{\it i} for the extended sample of
WTTS. The v~sin{\it i} were taken from the HBC, although for some stars we have
measured it in our spectra. These stars are Anon 1, IS Tau and 1E0255+2018; and
the measurements are, respectively, $<$15, 25\x10 y 30\x10~km s$^{-1}$.

The most striking feature of Fig. 9 is that there are no fast rotators
(v~sin{\it i}$\ge$40\kms) with low Li abundance. On the other hand, there is a
large dispersion of abundances among the slow rotators. The high Li abundances
in the fast rotators can be a real abundance effect or an artifact caused by
the presence of cool spots at the stellar surface. However, several works have
shown that there are no significant variations of the \lii\ doublet equivalent
width in spotted stars (Basri \et 1991, Mart\'\i n 1993a, Pallavicini \et
1993), and that the chemical abundances of elements other than lithium, i.e.
Ca, Fe, Ti, are not larger in fast rotating stars than in the slow rotators
(e.g. Balachandran \et 1988, Mart\'\i n 1993a). Thus, presently, there is no
observational evidence that high Li abundances in fast rotating TTS may be due
to spots.

The main source of uncertainty in  plots like Fig.~9. is the inclination angle
in the  v~sin{\it i}. In Fig.~10 we show the Li abundances of a limited sample
of WTTS and CTTS against the equatorial rotational velocity (deduced from
photometric periods, cf. Bouvier \et 1993a). We have restricted the sample to
K5-K7 single stars, corresponding to the theoretical mass range
0.9-0.5~M$_\odot$, to reduce possible variations of both lithium and rotation
with mass. In Table~9 we present the selected sample. The data on Li comes from
Basri \et1991, reanalyzed using our curves of growth (Fig.~6), and from this
work. The data on rotation comes from Bouvier \et 1993a.

The straight lines in Fig.~10 join points of different ages (3 and 10 Myr)
calculated by PKD for a mass of 0.8~M$_\odot$. Each line corresponds to a value
of initial angular momentum; the largest rotational velocities are for
J$_0$=1.6 x 10$^{50}$~g~cm$^2$ (model A2 of PKD), while the other two lines at
lower rotation velocities are for models A1 and A0 of PKD, J$_0$=5 x
10$^{49}$~g~cm$^2$, and 1.6 x 10$^{49}$~g~cm$^2$, respectively. All the CTTS in
Fig. 10, and all the WTTS, except for one, have Li abundances close to cosmic.
The single WTTS with low Li abundance (NTTS034903+2431) does not rotate very
fast, and presents less Li depletion than predicted by PKD. We believe that
Fig.~10 illustrates the point that before Li depletion takes place there is
already a wide angular momentum distribution, larger than one order of
magnitude. Thus, potentially, the initial angular momentum may be an important
factor in PMS Li depletion.

\begin{table*}
\caption[]{Lithium and Rotation in TTS of spectral types K5-K7}
\begin{flushleft} \begin {tabular}{lcccccc}
\hline Name  & Sp.T. & R/R$_\odot$ & v sin{\it i} & P$_{rot}$ & v$_{rot}$ & log
N(Li)  \\
& & & \kms & days & \kms & NLTE \\
\hline BP~Tau & K7 & 2.0 & $\le$10 & 7.6 & 11.8 & 3.2 \\
DK~Tau & K7 & 2.6 & 11.4 & 8.4  & 15.9 & 3.1 \\
GG~Tau & K7 & 2.6 & 10.2 & 10.3 & 12.9 & 3.3 \\
GM~Aur & K7 & 1.9 &12.4 & 12.0  & 8.0 & 2.7 \\
IW~Tau & K7 & 2.4 & $\le$9 & 5.6  & 22.0 & 2.8 \\
Lk~Ca~4 & K7 & 2.2 & 26.1 & 3.37  & 32.9 & 3.2 \\
Lk~Ca~7 & K7 & 2.0 & 13.0 & 5.64  & 18.0 & 3.2 \\
Lk~Ca~15 & K5 & 1.6 & 12.5 & 5.85  & 13.9 & 3.0 \\
Lk~Ca~19 & K5 & 2.2 & 18.6 & 2.24  & 47.6 & 3.0 \\
034903+2431 & K5 & 1.2 & 29 & 1.6  & 38.0 & 2.3 \\
042916+1751 & K7 & 1.7 &27 & 1.21  & 70.2 & 3.05 \\
V819~Tau & K7 & 1.8 & $\le$15 & 5.6 & 16.3 & 3.2 \\
V827~Tau & K7 & 2.2 & 18.5 & 3.75 & 29.6 & 3.2 \\
V830~Tau & K7 & 2.0 & 29.1 & 2.75 & 36.3 & 3.3 \\
V836~Tau & K7 & 1.6 & 29.1 & 7.0 & 11.7 & 3.1 \\
\hline
\end{tabular}
\end{flushleft}
\end{table*}

Our Fig. 9 recalls Fig. 3a of Balachandran \et 1988 for low-mass stars in
$\alpha$Per. These authors proposed two explanations:

\begin {enumerate}

\item Star formation over a time interval comparable with the timescale of Li
burning. Our data does not support this hypothesis because we see cases of Li
depletion among TTS younger than the $\alpha$Per members. Furthermore, there is
no apparent slow down trend with age in the PMS, but a bimodality of slow and
fast rotators that becomes more pronounced from the TTS to $\alpha$Per (Bouvier
\et 1993a).

\item Rapid rotational braking associated with enhanced Li depletion. Bouvier
\et 1993a found a systematic difference between the rotational velocities of
WTTS and CTTS. The WTTS are on the average fast rotators and could be the
progenitors of the very fast rotators in  $\alpha$Per and the Pleiades. On the
contrary, CTTS spin more slowly than WTTS, and could be the progenitors of slow
rotators in these clusters. Duncan 1993 and Soderblom \et 1993 have shown that
the rotational PMS evolution of solar-type stars can only be understood if the
angular momentum loss prior to settlement on the ZAMS is modest and rotation
independent. Consequently, a mechanism of angular momentum loss during PMS
evolution is not discarded by the previous works, but it is not required to
explain the observed distributions of rotational velocities in young low-mass
stars.

\end {enumerate}

The evolutionary models of PKD have predicted PMS Li depletion associated to
angular momentum loss. These authors could not explain the high Li abundances
in fast rotators of $\alpha$Per and the Pleiades, and advocated that they
resulted from late acretion along the Hayashi tracks. However,  this view is
not supported by the finding that CTTS rotate slowly. On the other hand, as we
mentioned before, PKD models for non rotating stars, and the models without
angular momentum redistribution predict Li depletion at 0.8~M$\odot$ roughly
consistent with the minima of Li abundances in slowly rotating WTTS. On the
contrary, models with angular momentum redistribution overestimate the Li
depletion at the same mass. Thus, there is no need to invoke a mechanism of PMS
Li depletion associated to angular momentum loss to explain the observed
lithium abundances in WTTS.

One possible drawback is if we could have missed any fast rotating TTS with low
Li abundance. The census of the fast rotating TTS is probably more complete
than that of the slow rotating  because fast rotators are more active and hence
easier to identify. If we have few fast rotators in our sample is because such
cases are rare among TTS. Duncan 1993 has not found any case with v~sin{\it
i}$>$50 \kms in a sample of 50 stars in the Orion Nebula region. We believe
that there is a real lack of fast rotating TTS with low Li abundances in the
range of masses and luminosities that we have studied. Our fast rotators are
not particularly young or massive, since they span the ranges of masses
1.0-0.7~M$_\odot$ and ages 10-50~Myr.

As we saw in Fig.~10,  TTS  in the mass range 0.9-0.5~M$_\odot$ show a wide
range of rotational velocities prior to Li burning . This initial spread in
rotation increases by about a factor 2 during PMS evolution (cf. Soderblom \et
1993). While slow-rotating WTTS destroy lithium at a rate consistent with
standard models without angular momentum redistribution,  fast-rotating WTTS
seem to inhibit Li depletion. If an inhibiting process takes place in fast
rotators during all PMS evolution, we could explain a large rotation-dependent
dispersion of Li abundances among the K-type stars of the ZAMS.

Members of the young open clusters $\alpha$Per (50~Myr) and the Pleiades
(100~Myr) offer the opportunity to study the final conditions of PMS evolution
for masses lower than  about 1.0~M$_\odot$. In Paper~II we will study the Li
abundances of low-mass stars in the Pleiades, and we will discuss further the
connection between PMS Li burning and rotation.

\section{Conclusions}

The main conclusions of this paper address the following points:

\begin{list}
 {--}{\setlength{\rightmargin}{\leftmargin}}

\item {\bf The initial lithium abundance:}

The statistical distribution of Li abundances in weak T Tauri stars has a
pronounced maximum at log~N(Li)=3.1, in close agreement with the cosmic lithium
abundance. We estimate that the initial Li abundance of TTS is remarkably
homogeneous, log~N(Li)$_{NLTE}$=3.11\x0.06 .

\item {\bf Pre-main sequence Li burning:}

There is clear evidence for PMS Li burning in our sample. As expected the
observations show that Li depletion increases towards lower luminosities. The
starting point occurs at L$_{*}$ between 0.9 and 0.4~L$_\odot$ depending on the
mass. At  L$_{*}\sim$~0.2~L$_\odot$ there is great Li depletion (over a factor
100) among stars with M$_{*} <$ 0.5~M$_\odot$. The observed luminosity of the
Li turnover at these masses is larger by a factor 4 than  theoretically
predicted.

Models without rotation can fit the maxima of Li depletion at masses around
0.8~M$_\odot$, but the same models overestimate the depletion at masses around
1.0~M$_\odot$. We find that PMS Li burning in stars with similar mass than the
Sun is less than a factor 2, implying that the solar Li abundance results from
evolution on the main sequence.

\item {\bf Lithium and rotation:}

At masses around 0.8~M$_\odot$ fast-rotating TTS do not show low Li abundances,
while slow-rotating TTS show decreasing Li abundances with luminosity. There is
no need for Li depletion associated to angular momentum loss since the observed
minimum Li abundances can be accounted for by non-rotating models. We interpret
the high Li abundances of the fast rotators as evidence for an inhibiting
process of Li depletion associated to rapid rotation.

We have shown that there is a wide range of rotation velocities in TTS with
masses 0.9-0.5~M$_\odot$ before Li depletion starts to be noticed. Thus,
differences in initial angular momentum and inhibition of Li depletion  in fast
rotators can produce a spread of Li abundances on the ZAMS for stars of the
same mass ($\sim$0.8~M$_\odot$). \end{list}

\acknowledgements {We acknowledge the help of
Ram\'on J. Garc\'\i a L\'opez and John K. Webb to collect some of
the data presented in this paper. We are also grateful to
Franca D'Antona for sending us the tracks in her work with Italo Mazzitelli
prior to publication. This work was partially supported by the Spanish
DGICYT under projects PB89-0375-C02-01 and PB91-0526}

\newpage

{\it Figure captions:}

{\bf Figure 1:} T Tauri spectra taken with the ISIS spectrograph
at the William Herschel
telescope. The position of the \li\ feature is marked. All the continua
have been normalized and displaced by a constant factor.

{\bf Figure 2:} T Tauri spectra taken with the intermediate dispersion
spectrograph (IDS) at the Isaac Newton telescope.
Dispersion = 0.036~nm pix$^{-1}$.

{\bf Figure 3:} T Tauri spectra taken with IDS at the Isaac Newton telescope.
Dispersion = 0.036~nm pix$^{-1}$.

{\bf Figure 4:} T Tauri spectra taken with IDS at the Isaac Newton telescope.
Dispersion = 0.022~nm pix$^{-1}$.

{\bf Figure 5:} Our sample of weak T Tauri stars in the H-R diagram.
Evolutionary tracks and isochrones are from set~1 of D'Antona \& Mazzitelli
1993.

{\bf Figure 6:} Curves of growth for the \lii\ resonance doublet.
{}From above downwards the pairs of curves  (one solid for LTE and one dashed
for NLTE) have been calculated for T$_{\rm eff}$ of
3500, 4000, 4500, 5000 and 5500~K, respectively. In all computations we
used  log~g=4, microturbulence of 2~\kms and solar metallicity.

{\bf Figure 7:} Statistical distribution of lithium abundances in our sample
of weak T Tauri stars.

{\bf Figure 8:} Lithium abundances in WTTS plotted against their
bolometric luminosities. Superimposed are theoretical curves from
 set~1 of D'Antona \& Mazzitelli 1993 for 0.8 (solid) and
0.4~M$_\odot$ (dots and dashes), and from Table~1 of Pinsonneault \et1990
for the same masses (see text).

{\bf Figure 9:} Lithium abundances in WTTS plotted against their projected
rotational velocities. Open squares indicates measurements of  vsin{\it i},
whereas open triangles indicate upper limits on  vsin{\it i}.

{\bf Figure 10:} Lithium abundances in WTTS (crosses) and CTTS (open squares)
of spectral types K5-K7 plotted against their equatorial rotational velocities.
The straight lines join points of ages 3 and 10~Myr, calculated by Pinsonneault
\et1990 for 3 different values of initial angular momentum
(models A0, A1 and A2), at the same mass of 0.8~M$_\odot$.

\end{document}